# Demonstration of white light cavity effect using stimulated Brillouin scattering in a fiber loop

H. N. Yum, J. Scheuer, M. Salit, P. R. Hemmer and M. S. Shahriar

*Abstract*—A passive white light cavity (WLC) based on a fiber resonator can be used for high-bandwidth optical data buffering. Here, we report on experimental studies of such a WLC, employing stimulated Brillouin scattering (SBS) for producing the negative dispersion, using two different configurations. In one configuration, an absorption peak produced by a Brillouin pump is used. In the other configuration, two gain peaks produced by two separate Brillouin pumps are employed. In each case, we see evidence of the WLC effect. However, the range of parameters accessible experimentally limits the degree of the WLC effect significantly. We present a theoretical analysis for the optimal combinations of parameters, such as a high Brillouin gain coefficient and a low transmission loss, necessary for achieving the condition of a vanishing group index, as required for creating an ideal WLC.

## I. Introduction

A White light cavity (WLC) is a unique type of resonator which is designed to resonate over a continuous range of frequencies. This is in contrast to conventional cavities which resonate at discrete frequencies determined by their optical roundtrip length. The WLC concept exhibits unique properties which render it scientifically interesting as well as attractive for various applications. One of the most important properties of WLC is the elimination of the traditional relation between the linewidth of the cavity and its quality factor. In contrast to conventional cavities, a WLC possesses a broader linewidth than that of a conventional cavity with the same finesse [1, 2]. In addition, the sensitivity of the lasing frequency of a WLC based laser to changes in the cavity length is substantially higher than that of conventional cavity based lasers. The combination of wide bandwidth on one hand and large finesse on the other is highly attractive for various applications, primarily in sensing and telecommunication. WLC based schemes have been proposed for enhanced gravitational waves detection [3, 4], sensing [5, 6] and for trap-door data buffering with a large delay-bandwidth product [7].

The key element in the realization of a WLC is a dispersive phase compensation mechanism having a negative phase slope with respect to the frequency (larger phase shift for lower frequencies) [6]. This mechanism equates the phase accumulated in a roundtrip by each frequency component, setting it to a multiple integer of $2\pi$. Various approaches have been proposed and studied for the realization of such phase compensation mechanism, such as dual-pump Raman gain profile in Rb vapor cell [7, 1], four wave mixing [8], and an intra-cavity resonator [6, 9].

In this paper, we explore theoretically and experimentally the feasibility of the realization of fiber optics based WLC utilizing Brillouin scattering (absorption and gain) for attaining the necessary negative group delay. In Section II, we present the concept and the theoretical model. In Section III, we study the realization of a WLC utilizing Brillouin absorption induced by a single pump. In Section III, we study the realization of a WLC utilizing a notch in a Brillouin gain profile induced two spectrally separated pumps. In Section V, we study the dependence of the attained cavity linewidth on the Brillouin gain/absorption conditions. In Section VI, we study the parameter constraints that limit the performance of the WLC, and suggest approaches for overcoming these limitations. Finally, in Section VIII, we summarize the results and present an outlook for future studies.

## II. Cavity concept and Theoretical Model:

Consider the fiber ring cavity in an all-pass filter configuration as illustrated in Fig. 1. The cavity consists of a 2X2 coupler and a fiber loop. Assuming the coupler lossless, its transfer matrix (the relation between the inputs and the outputs fields) is given by:

$$\begin{pmatrix} b_1 \\ b_2 \end{pmatrix} = \begin{pmatrix} \sqrt{1-K} & j\sqrt{K} \\ j\sqrt{K} & \sqrt{1-K} \end{pmatrix} \begin{pmatrix} a_1 \\ a_2 \end{pmatrix} \qquad (1)$$

This work was supported in part by AFOSR under grant number FA9550-10-1-0228, and by the Israeli Science foundation.

H. N. Yum was with Northwestern University, Evanston, IL 60208 USA. He is now with Research Laboratory of Electronics at Massachusetts Inst. of technology, Cambridge, MA 02139 USA (e-mail: hn_yum@yahoo.com).

J. Scheuer is with the Department of Electrical Engineering, Northwestern University, Evanston, IL 60208 USA on leave from the school of Electrical Engineering, Tel-Aviv University, Tel-Aviv 69978, Israel (e-mail: kobys@eng.tau.ac.il).

M. Salit and M. S. Shariar, are with the Department of Electrical Engineering, Northwesten University, Evanston, IL 60208 USA

P. R. Hemmer is with the faculty of Electrical Engineering and Computer Science, Texam A&M University, College Station, TX 77843-3126 USA.

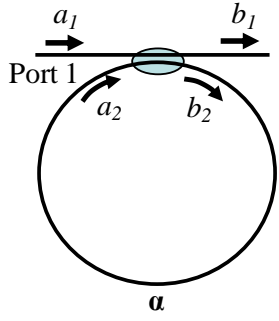

Fig. 1. Schematics of a fiber ring resonator and definitions of the fields in the cavity.

where $K$ is the power coupling coefficient. The relation between $a_2$ and $b_2$ is determined by the fiber loop transmission gain/loss, $\alpha$, and the roundtrip phase shift, $\theta$:

$$a_2 = \alpha \exp(j\theta) \cdot b_2 \qquad (2)$$

Equations (1) and (2), provide the intensities inside the cavity, $|b_2|^2$ and the transmitted power $|b_1|^2$:

$$\left|\frac{b_1}{a_1}\right|^2 = \frac{\alpha^2 + |t|^2 - 2\alpha|t|\cos\theta}{1 + \alpha^2|t|^2 - 2\alpha|t|\cos\theta}$$
$$\left|\frac{b_2}{a_1}\right|^2 = \frac{1 - |t|^2}{1 + \alpha^2|t|^2 - 2\alpha|t|\cos\theta} \qquad (3)$$

where $t = \sqrt{1-K}$. Through the rest of the paper we assume that the input intensity $|a_1|^2 = 1$. The roundtrip phase, $\theta$, determines the resonant properties of the cavity such the linewidth, the free spectral range (FSR), etc. In conventional fiber loop, this phase shift is essentially given by $k \cdot L$ where $L$ is the length of the loop and $k$ is the wavenumber given by $k = n(\omega)\omega/c$. $n$, $\omega$ and $c$ are respectively the refractive index, the angular frequency and the speed of light and the fiber loss can be neglected in the settings discussed in this paper. Modifying the spectral properties of $\theta$ requires the modification of the refractive index $n(\omega)$. Here, we modify $n$ using Brillouin gain/absorption, exploiting the relations between the real and imaginary parts of the linear electric susceptibility (Kramers-Kronig relations relations).

Specifically, we are interested in modifying the roundtrip phase in order to realize a WLC, which is characterized by vanishing group index, i.e. $n_g=0$. Because the conventional propagation in the fiber loop provides a positive group index, it is necessary to compensate it by inducing a negative contribution to the group index through the Brillouin gain/absorption. Attaining this requires a notch in the spectral transmission induced by the Brillouin scattering process. In the sections below we explore two approaches for attaining such transmission notch using Brillouin scattering: Brillouin absorption (section III) and spectrally separated dual Brillouin gains (section IV).

Attaining Brillouin gain/absorption in the fiber loop requires the launch of a pump beam into the input port of the cavity (Fig. 1). The effectiveness of the Brillouin scattering process depends on the intensity of the pump and interaction length between the pump and the signal where higher pump intensities and longer interaction lengths increase the gain/absorption. To increase the effectively of the process it is advantageous to set the wavelength of the pump to coincide with one of the resonance frequencies of the cavity. As a result, the intensity of the pump inside the cavity is given by $I_{pump}^{cav} = I_{pump}^{in}(1-t^2)/(1-\alpha t)^2$, where $I_{pump}^{in}$ is the intensity of the pump which is launched in port 1. Depending on $t$ and $\alpha$, $I_{pump}^{cav}$ could be substantially larger than $I_{pump}^{in}$.

The pump beam in the fiber loop induces Brillouin absorption (gain) for contra-propagating probe beam having frequency which is higher (lower) than that of the pump by the Brillouin frequency shift $\Delta\nu_B$. Referring to the single frequency pump case, the Brillouin absorption coefficient and the corresponding modification to the wavenumber are given by:

$$\alpha_{Br} = \frac{-\frac{1}{2}g_0 I_P}{1+4(\nu-\nu_B)^2/\Gamma_B^2} \;;\; \beta_{Br} = \frac{-g_0 I_P(\nu-\nu_B)/\Gamma_B}{1+4(\nu-\nu_B)^2/\Gamma_B^2} \qquad (4)$$

where $\nu_B$, $g_0$, $\Gamma_B$ and $I_P$ are respectively the Brillouin maximal absorption frequency, gain coefficient, linewidth, and pump intensities per unit area. In the presence of the pump the roundtrip phase shift of the probe can be expressed as $\theta = k \cdot L = n_f \omega L/c + 2\pi\beta L$, where $n_f$ and $c$ are the mean index of the fiber and $k$ represent the modified and frequency dependent wavenumber of the probe. Achieving the WLC condition at the center of the transmission notch requires the pump intensity to satisfy:

$$n_g = n_f - c \cdot g_0 I_{pump}^{cav}/\Gamma_B = 0 \qquad (5)$$

Next, consider $\theta$ in the presence of bi-frequency Brillouin pump. The pumps (and correspondingly the peak gains) are separated by $2\Delta$ in frequency in order to form an effective transmission notch at the center of the combined gain profile. Note that in order to exploit the Brillouin gain, the probe frequency must be lower than that of the pumps. The Brillouin gains $\alpha_{Br}$ and the phase $\beta$ in this case are given by:

$$\alpha_{Br} = \frac{\frac{1}{2}g_0 I_{P1}}{1+4(\nu-\nu_B-\Delta)^2/\Gamma_B^2} + \frac{\frac{1}{2}g_0 I_{P2}}{1+4(\nu-\nu_B+\Delta)^2/\Gamma_B^2} \qquad (6a)$$
$$\beta_{Br} = \frac{g_0 I_{P1}(\nu-\nu_B-\Delta)/\Gamma_B}{1+4(\nu-\nu_B-\Delta)^2/\Gamma_B^2} + \frac{g_0 I_{P2}(\nu-\nu_B+\Delta)/\Gamma_B}{1+4(\nu-\nu_B+\Delta)^2/\Gamma_B^2} \qquad (6b)$$

where $I_{P1}$ and $I_{P2}$ are pump intensities per unit area. For simplicity, we consider equal pump intensities such that $I_{P1} = I_{P2} = I_P$ (W/m$^2$). Achieving the WLC condition $\nu=\nu_B$ requires the intensities and frequency separation of the two pumps to satisfy:

$$n_g = n_f - 2c \cdot g_0 I_{pump}^{cav}/\Gamma_B \cdot \frac{1-(2\Delta/\Gamma_B)^2}{[1+(2\Delta/\Gamma_B)^2]^2} = 0 \qquad (7)$$

Finally, it is instructive to estimate the impact of the group index on the cavity linewidth as it is convenient to estimate the group index by measuring the cavity linewidth. The width of the notch in Eq. (3a) is determined essentially by the roundtrip phase, $\theta$, which can be written as [10]:

$$\theta = 2\pi\Delta\omega/\Delta\omega_{FSR} \qquad (8)$$

where $\Delta\omega$ and $\Delta\omega_{FSR}$ are respectively the angular frequency shift from the resonance frequency and the free spectral range. The free spectral range of the cavity is determined by the group index and the cavity length [10]:

$$\Delta\omega_{FSR} = 2\pi c/n_g L \qquad (9)$$

Let us define the cavity bandwidth as the frequency shift for which $\theta$ reaches a certain value - $\theta_{BW}$. Note that as long as $\alpha$ and $t$ are fixed, the cavity lineshape is determined solely by the roundtrip phase and the linewidth is determined according to $\theta(\Delta\omega_{BW}/2) = \theta_{BW}$, i.e.:

$$\Delta\omega_{BW} = 2\theta_{BW} c/n_g L \qquad (10)$$

Defining $\gamma_0$ as the cavity linewidth in the absence of the Brillouin pumps induced dispersion (i.e. $n_g=n_f$), yields a simple relation between the group index and the cavity linewidth - $\Delta\omega_{BW} = \gamma_0 n_f/n_g$. Clearly, this expression is valid only if the group index is the dominant factor determining the cavity linewidth. Close to the WLC condition, when the group index approaching zero, higher dispersion order become non-negligible and form the dominant factor determining the cavity lineshape and linewidth.

### III. WLC BASED ON BRILLOUIN ABSORPTION

We first attempt to realize a fiber-based WLC by utilizing Brillouin absorption for attaining negative dispersion. The experimental setup is depicted in Fig. 2. The cavity consists of a 12m long single mode fiber ring. A section of the fiber cavity is wound around a piezo-electric transducer (PZT) tube. By applying voltage to the PZT tube the cavity length can be modified. Two fiber polarization controllers (FPCs) are incorporated in the cavity in order to control the polarization of the light circulating in the cavity. The precise cavity length is adjusted so that Brillouin frequency shift $\Delta\nu_B$, measured to be ~10.867 GHz, is an integer multiple of the free spectral range (FSR). This is in order to simultaneously attain resonant pumping, and generating the Brillouin absorption line at one of the cavity resonance frequencies. The resulting FSR of the cavity is ~17.32 MHz. The pump and the probe source is a laser diode (LD) which is stabilized by a fiber loop mirror [11] utilizing 10% of the LD output. The remaining 90% is split by a 50:50 polarization maintaining coupler (PMC) where one the output serves for the probe and the other for the pump. The pump signal passes through an electro-optic modulator (EOM-1), and amplified by an Erbium-doped fiber amplifier (EDFA). For the case discussed in this section, EOM-1 is not activated, and the pump signal retains the LD frequency. The pump output (port 3 of the cavity coupler) is detected and fed to phased locked loop (PLL) employing a lock-in-amplifier (LIA-1) and an AC servo which modulates the cavity length by applying a signal to the PZT tube. The feedback loop is used for locking the ring cavity to the pump frequency.

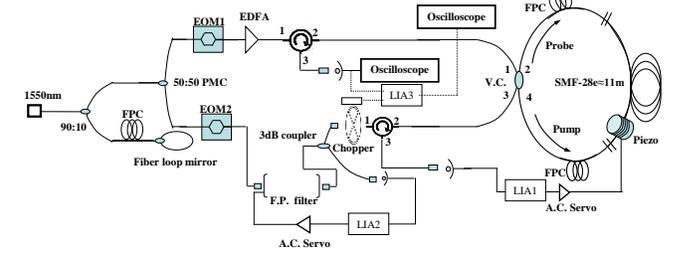

Fig. 2. Schematics of the experimental setup to observe the WLC effect. The components are: LIA, lock-in amplifier; EOM, electro-optic modulator; FPC, fiber polarization coupler; PMC, polarization maintaining coupler; and VC, variable coupler. See text for additional details.

The second output of the PMC is modulated by an EOM (EOM-2) at the Brillouin frequency, $\Delta\nu_B$, thus generating two sidebands with frequencies that fall within the Brillouin gain/absorption lines. The fundamental frequency is suppressed by adjusting the DC voltage applied to EOM-2. The two sidebands are filtered by a Fabry-Perot spectral filter (FPSF), which is controlled by another LIA (LIA-2). The output of the FPSF is equally split by a 3dB coupler where one of the outputs is sent to a detector. The detected signal is demodulated by LIA-2, and is utilized for locking the FPSF to the transmission peak one of the two EOM sidebands. For the study discussed in this section, which utilizes Brillouin induced absorption, the FPSF is locked to transmit the upper sideband at frequency $\nu_L + \Delta\nu_B$. The second output of the 3dB coupler (which is also at frequency $\nu_L + \Delta\nu_B$) serves as the probe signal for the ring cavity. The modulation frequency of the probe is slowly swept around $\Delta\nu_B$ (while the FPSF tracks the upper sideband) by approximately $\Delta\nu=3$MHz.

In order to enhance the signal to noise ratio, the probe is modulated by a mechanical chopper before it is injected to the ring cavity. The probe signal at the cavity output is detected and demodulated by a third LIA (LIA-3), which is synchronized with the chopper. Fig. 3 depicts experimental results (a) and numerical simulations (b) of the cavity transfer function in the vicinity of the Brillouin absorption line as a function of the pump power. To obtain the probe transmission dip (black) in the absence of SBS, probe was modulated at 7.0145 GHz which is sufficiently remote from the Brillouin frequency shift. As a result the probe does not interact with the Brillouin gain/absorption induced by pump which is being used for locking the cavity. The linewidth of the ordinary cavity is measured to be 0.24 MHz. Next, we modulate the probe at the Brillouin frequency shift. The FPSF is locked to the upper (lower) sideband for producing SBS absorption (gain) at the probe frequency. Note that for Brillouin gain studies, the lower sideband of the modulated probe was employed. The pump power maintained below the Brillouin lasing threshold to avoid interference between the probe and the Brillouin lasing fields. Fig. 3(a) depicts the experimental characterization of the cavity lineshape for various pumping conditions. For simplicity, we define $G \equiv -\frac{1}{2}g_0 I_p$ for absorption, and $G \equiv \frac{1}{2}g_0 I_p$ for gain. Fig. 3(b) depicts simulations of the cavity lineshape associated with the experimental results. The depth of the notch at the cavity resonance is given by Eq. 3a - $|b_1/a_1|^2 = (\alpha-t)^2/(1-\alpha t)^2$. This expression also applies when the cavity is pumped with $\alpha$ replaced by $\alpha \exp(\alpha_{Br} L)$. Fig. 3(a) shows that the cavity is over-coupled ($\alpha > |t|$) in the absence of Brillouin pump; critically coupled ($\alpha \exp(\alpha_{Br} L) = t$) when additional cavity loss is introduced by SBS absorption induced by 2mW pump power; under-coupled ($\alpha \exp(\alpha_{Br} L) < t$) as the SBS absorption is increased; and over-coupled ($\alpha \exp(\alpha_{Br} L) > t$) when SBS gain is introduced.

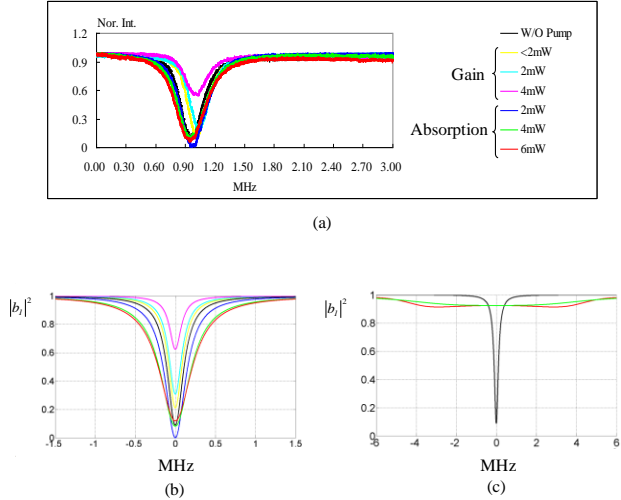

Fig. 3. (a) Cavity response measured for SBS absorption and gain by varying the pump power. (b) Numerical simulations for the cavity response corresponding to the experimental results. The colors of the graphs are the same as in (a). $G$ values used for the simulation are: $G$=0.0002 for $n_g$ =1.4538 (yellow); $G$=0.0003 for $n_g$ =1.4557 (cyan); $G$=0.0005 for $n_g$=1.4595 (magenta); $G$=−0.0005 for $n_g$ =1.4405 (blue); $G$=−0.0015 for $n_g$=1.4214 (green); $G$=−0.0018 for $n_g$=1.4156 (red). The negative (positive) $G$ corresponds to SBS absorption (gain). (c) Calculated cavity response: (i) without SBS gain or absorption (black); (ii) on the WLC condition corresponding to $n_g$=0, $G$=0.0759 (red); (iii) non-dispersive cavity (i.e. $n_g$=$n_f$, $n_f$ =1.45 for SMF-28e) including only SBS absorption-induced cavity loss.

Based on the experimental result it is possible to find the cavity parameters – $\alpha^2$=0.97 and $|t|^2$=0.945. Comparing the experimentally obtained cavity lineshape with that attained by the model, allows us to determine the value $|G|$ corresponding to each pump level (see Fig. 3). We note that the values of $|G|$ for SBS gain and absorption scenarios should in principle be identical for the same pump level. This, however, that does not agree with the experimental results. For example, at a pump level of 2mW we find $|G|$=5x10$^{-4}$ for SBS gain and $|G|$=1.5x10$^{-4}$ for SBS absorption. While the origin of this discrepancy is not yet clear, it may stem from the difference in the accuracy in which the Brillouin pump is locked to the cavity resonance for the gain and absorption cases.

Generally speaking, there are two factors that contribute to the linewidth broadening: (i) the additional cavity roundtrip loss induced by the SBS absorption and (ii) the modified roundtrip phase response (i.e. the "WLC effect"). Although the two mechanisms cannot be separated in practice, their individual contribution to the broadening can be evaluated [12].

In order to determine whether the WLC effect is distinguishable from the broadening due to the SBS absorption, it is instructive to plot the cavity lineshape for two cases: (1) An ideal WLC realized by employing only the dispersion induced by SBS (setting the pump to attain $n_g$=0) but neglecting the accompanied loss and (2) accounting only for the Brillouin induced loss (at the same pumping conditions) but neglecting the accompanied dispersion. Fig. 3(c) depicts a comparison between the cavity linesahpes in both cases. Clearly, the difference between the linewidths is very small, thus indicating that the contribution of the SBS-induced loss is as important as the SBS induced phase shift. Therefore, we conclude that at least for the case of below lasing threshold pump levels, the WLC effect is not clearly distinguishable from the SBS absorption-induced broadening.

## IV. WLC BASED ON DUAL BRILLOUIN GAIN PROFILES

Following the indefinite results discussed in section III, we study an alternative approach for realizing a WLC by employing spectrally separated dual SBS gain profiles. A schematic of the experimental setup is shown in Fig. 4. The experimental procedure is similar to that described in section III except that the pump is modulated by EOM-1 at 17.32MHz, corresponding to the FSR of the fiber cavity, to provide the dual Brillouin pumps. The fundamental lobe was suppressed. The modulation frequency of the pump was chosen in order to attain resonant Brillouin pumping at both frequencies and set the notch between the Brillouin gain profiles at one of the cavity resonance frequencies. The probe was modulated at the Brillouin frequency shift (10.867GHz) where the fundamental was suppressed as well. Although only the lower frequency sideband of the probe, experiencing the Brillouin gain, is intended to serve as the probe, the upper sideband was not filtered out as it was expected to be attenuated significantly by the absorption induced by the pumps. However, as shown below, this assumption was found to be too optimistic and in fact, the absorption experienced by this sideband is relatively modest because it is located between the two absorption dips. As a result, the impact this sideband on the overall transmission of the probe is non-negligible. In the experiment described in section V below this drawback of the experimental setup is mitigated.

Fig. 5(a) illustrates the properties of the pump and the Brillouin gain spectra with respect to the cavity modes. The dual-frequency pump and the corresponding dual Brillouin gains are resonant with cavity modes. The non-pumped cavity is set to be critically coupled ($t=\alpha$) by adjusting the coupling ratio of the variable coupler (VC) [[i]3]. At this condition, the full width at half maximum (FWHM) was measured to be 0.36MHz, corresponding to $\alpha^2=t^2$=0.93. Fig. 5(b) depicts the experimentally measured cavity transmission at various Brillouin pump levels. When the cavity is pumped at 2mW, the cavity remains very close to the critical-coupling condition, and becomes more and more over-coupled as the pump power is increased. Fig. 5(c) shows the theoretical calculation of the signal spectrum at the cavity output, were both sidebands of the signal have been taken into account. The wide "basin" of the spectrum stems from the absorption of the higher frequency sideband while the narrower peak at the center is the amplified lower frequency sideband of the probe which experiences Brillouin gain in the cavity.

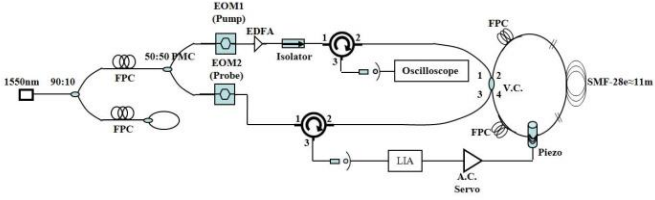

Fig. 4. Schematics of the experimental setup for dual SBS gains WLC.

Note that although the relatively flat spectral transfer profiles measured for the three highest pump powers (8.5 mW, 9.2 mW, and 9.6 mW) may seem like a WLC behavior, the theoretical study indicates that the flat profile is caused by the balancing of the effects (gain and loss) experienced by the two probe sidebands.

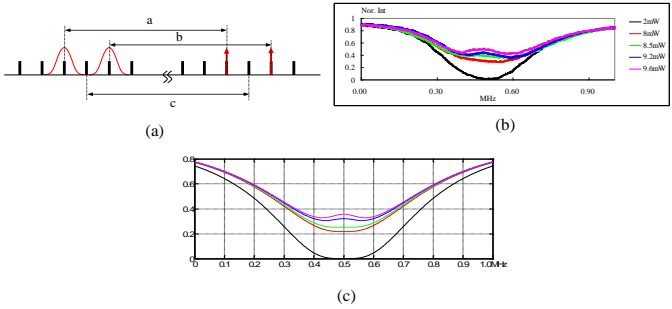

Fig. 5. (a) Cavity modes (black bars), Brillouin pump spectra (red arrows) and gain lines (Lorentzians in red). The separations represented by a, b and c are equal to $\Delta \nu_B$=10.867GHz. The cavity FSR is 17.32MHz. (b) Probe transmission as its frequency is scanned between the two gain peaks for 5 different pump power levels. The gain separation is kept fixed at 2×FSR=34.64MHz. (c) Numerical simulations performed with the following parameters, accounting for the gain (absorption) experienced by the lower (upper) sideband of the probe: $G$=0.0055 ($n_g$=1.4432) for 2mW, $G$=0.0223 ($n_g$=1.4223) for 8mW, $G$=0.0229($n_g$=1.4215) for 8.5mW, $G$=0.0238 ($n_g$=1.4204) for 9.2mW, and $G$=0.0242 ($n_g$=1.4199) for 9.6mW. The proportionality constant between the G values and the pump power levels is chosen for obtaining the best fit to the experimental results, and is also in close agreement (within 10%) with the estimated value obtained from the expression $G \cong g_0 I_P / 2$, using $g_0 = 1.01 \times 10^{-11}$ m/W, cavity build-up factor of ~14, and an effective core diameter of ~8.2 μm. The group indices are calculated from the G values.

## V. WLC BASED ON DUAL BRILLOUIN GAIN PROFILES - IMPACT OF THE PUMPS SPECTRAL SEPARATION

The gain and dispersion profiles generated by the dual Brillouin pump induce two processes which simultaneously broaden and narrow the transmission linewidth of the cavity. The negative dispersion broadens the linewidth while the residual gain narrows it. In order to reduce the narrowing effect, we increased the gain separation to four times the cavity free spectral range as illustrated in Fig. 6(a). In these conditions, the overlap between the tails of the Brillouin gains is substantially smaller and the corresponding residual gain at the center is very small.

The experimental study described in section IV revealed that it is important to filter out the higher frequency sideband of the probe (experiencing Brillouin absorption) because its impact on the measured linewidth is non-negligible. Thus, the experiment was repeated where upper sideband ($\nu_L+\Delta \nu_B$) filtered out by the FPSF as described in Section III. Figure 6 depicts experimentally measured and theoretically calculated cavity linewidth for various pump powers. At pump level of 4mW, the broadening and narrowing effects are balanced and the resulting linewidth is essentially identical to the reference linewidth measured without pumps. As the pump power is increased, the narrowing effect becomes dominant yielding a linewidth which is narrower than that of the "cold" cavity. Figure 6(c) shows theoretical calculations of the expected linewidth for the pump levels of Fig. 6(b), exhibiting good agreement with the experimental data.

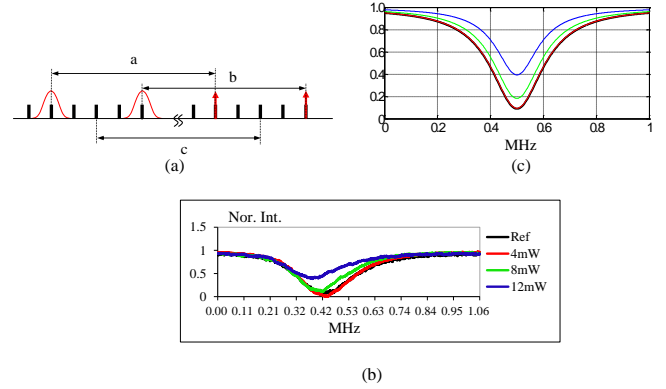

Fig. 6. (a) Cavity modes (black bars), bi-frequency Brillouin pump spectra (red arrows) and gains (Lorentzians in red). The separations a, b and c correspond to $\nu_B$ (=10.867GHz). (b) Experimental measurement of the cavity response for dual SBS gains. (c) Numerical simulations with parameters: $G$=0.5×10$^{-3}$, $n_g$=1.4498 (4mW), and $G$=0.8×10$^{-2}$, $n_g$=1.447 (8mW), and $G$=1.8×10$^{-2}$, $n_g$=1.4433 (12mW). The colors correspond to those of panel (b).

## VI. PRACTICAL REALIZATION OF WLC – WHAT DOES IT TAKE?

Following the experimental studies described in the previous sections, we discuss here the parameter constraints and requirements for the realization of a WLC using dual Brillouin gain. The outcome of this discussion is a proposal of a cavity system which satisfies these constraints.

We consider the cavity parameters - *FSR*, roundtrip loss and Brillouin gain needed to attain a desired group index. For illustration purposes we target three values of group index - $n_g$=0, 0.5 and 1.1, where we expect observe cavity linewidth broadening. $n_g$=0 corresponds to the ultimate WLC which ideally yields infinite linewidth but in practice, the linewidth is limited by higher order dispersion. According to Eq. (10), the cavity linewidth is expected to be broadened by factors of 3 and 1.3 for $n_g$=0.5 and $n_g$=1.1 respectively.

One of the important constraints on the Brillouin gain is that the cavity should be kept below lasing threshold, not only for the cavity mode of interest (the one located between the Brillouin gain peaks) but also for all other cavity modes. Thus, in what follows we set the cavity FSR such that $\Delta \nu_{FSR} \cdot \left(N+\frac{1}{2}\right) = \Delta \nu_B$ and the pumps separation to satisfy $2\Delta = \Delta \nu_{FSR}$. By doing so, we ensure that the cavity mode located between the gain peaks experiences more Brillouin gain than all other cavity modes and that if this mode is kept below threshold then no other cavity mode would lase. The lasing threshold for a conventional cavity is $a = e^{-\alpha_{loss}L}$ where $a$ is the gain, $\alpha_{loss}$ is the propagation loss in the fiber [14] and *L* is the cavity length. The gain coefficient induced by a dual Brillouin pump in the middle of the individual gain profiles is derived from Eq. (4) - $a_{Br} = 2G/(1 + 4\Delta^2/\Gamma_B^2)$. The roundtrip propagation gain/loss of the probe at this frequency is, therefore, $a \cdot e^{a_{Br}L_{eff}}$ where $L_{eff} = -L(1-a)/\ln a$ is the effective length [14]. To eliminate Brillouin lasing, the overall roundtrip gain (including the coupling loss) must be smaller than unity, i.e. $a_{Br}L_{eff} < -\ln ta$. Starting with the desired group index we use to express the *G* in terms of $n_g$ using Eq. (7):

$$G = (n_f - n_g) \cdot \Gamma_B/4c \cdot \frac{[1+(2\Delta/\Gamma_B)^2]^2}{1-(2\Delta/\Gamma_B)^2} \quad (11)$$

From *G*, we can immediately deduce the Brillouin gain coefficient - $a_{Br}L_{eff}$. Figure 7 depicts the dependence of $a_{Br}L_{eff}$ on the desired group index and the cavity FSR for various fiber propagation loss values. The lasing threshold, $-\ln ta$, is superimposed. The rest of the parameters are $t$=0.955, $\Gamma_B$=10MHz, and $n_f$=1.45, and the gain separation is the cavity FSR i.e. $2\Delta = \Delta v_{FSR}$. According to (11), reducing the group index necessitates larger *G* and correspondingly larger $a_{Br}L_{eff}$. Given the desired group index, Fig. 7 allows us to determine the parameter space (FSR and gain) in which the system can operate without lasing. For example, assuming the propagation loss in the cavity is *a*=0.99 then in order to attain zero group index the cavity FSR must exceed 48MHz is order to suppress lasing.

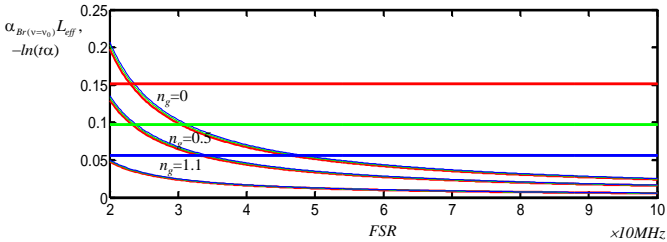

Fig. 7. Curves represent $\alpha_{Br(v=v_0)}L_{eff}$ obtained for given $n_g$. Straight lines represent $-\ln(t\alpha)$. Red: $\alpha$ =0.9, green: 0.95, blue: 0.99. The *FSR*s which satisfy $\alpha_{Br(v=v_0)}L_{eff} < -\ln(\alpha)$ are: (i) for $n_g$=0, *FSR*>23MHz with $\alpha$=0.9, *FSR*>31MHz with $\alpha$=0.95, *FSR*>48MHz with $\alpha$=0.99, (ii) for $n_g$=0.5, *FSR*>18MHz with $\alpha$=0.9, *FSR*>23MHz with $\alpha$=0.95 and *FSR*>34MHz with $\alpha$=0.99 (iii) for $n_g$=1.1, *FSR*>13MHz with $\alpha$=0.9, *FSR*>15MHz with $\alpha$=0.95 and *FSR*>19MHz with $\alpha$=0.99. *FSR* lower than 20MHz is not shown in the graph.

Fig. 8 depicts the value of *G* needed to generate three levels of group as a function of the cavity FSR (again, the pump separation is kept equal to the FSR). Clearly, the necessary gain to attain a desired group index increases for larger FSR due to the larger separation between the peaks of the induced gain profiles.

It is instructive to relate the values of G plotted in Fig. 8 to a physical property such as the corresponding input pump power $I_j$. The input pump power is related to field intensity circulating inside the cavity, $I_p$, which on resonance satisfies $I_j = I_p A_{eff}/\eta$ where $A_{eff}$ is the effective area of the fiber and $\eta$ is the cavity build up factor, $\eta = (1-t^2)/(1-\alpha t)^2$. For single mode fibers, $A_{eff}$= 5×10⁻¹¹ m² [i] and $g_0 \approx 10^{-11} m/W$. Figure 9 depicts the necessary input power needed to attain a desired group index given the fiber losses and the cavity FSR. It must be emphasized that in order to avoid lasing the FSR of the cavity must exceed a certain value. For example, attaining $n_g$=0 with $\alpha$=0.9 requires the FSR to be larger than 23MHz. This, in turn, sets a lower limit on the necessary pump power range which in this case must satisfy $I_j > 0.62W$. In practice, gain saturation effects may limit *G* and prevent reaching group index change which is sufficient for observing the WLC effect, in particular if short cavities having large FSR are utilized. Therefore, it may be beneficial to use optical fibers possessing high Brillouin coefficient and small effective area, rather than conventional silica fibers. For example, tellurite fiber ($g_0$=1.6986×10⁻¹⁰ m/W, $A_{eff}$= 0.6967×10⁻¹¹ m²) [15, 16] or chalchogenide glass fiber ($g_0$=6.08×10⁻⁹ m/W, $A_{eff}$= 3.9×10⁻¹¹ m²) [17, 18] are promising candidates. However, these fibers exhibit relatively high transmission loss which may attenuate he pump substantially even along a single round-trip, rendering the resonant pumping scheme less attractive. For such fibers, a non-resonant pumping scheme as depicted in Fig. 10 may be preferable. Figure 10 depicts a cavity comprising high Brillouin coefficient fibers and a low loss optical circulator. Because the Brillouin amplified signal circulates in an opposite direction to the pump, it experiences a closed cavity while the pump exits the cavity after a single roundtrip. Another attractive possibility to overcome the pump attenuation is to incorporate an optical amplifier in the cavity.

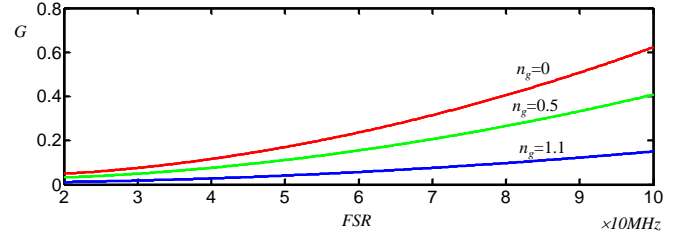

Fig. 8. *G* dependence on the *FSR* for different group indexes

VII. CONCLUSION

We studied two configurations for realizing WLC using SBS in fibers. The first configuration utilizes the absorption induced by a resonantly enhanced, single frequency, Brillouin pump. The cavity length is set to ensure that the absorption line coincides with one of the cavity modes. We found that the transfer function of the cavity in the vicinity of this mode exhibits a broadened linewidth. Theoretical studies confirm that the observed linewidth broadening stems from a combination of the WLC effect, i.e. the dispersion induced by the SBS, and excess loss induced by the SBS process. The analysis also reveals that this is a fundamental constraint for Brillouin absorption induced WLC. The first configuration utilizes SBS gain induced by pumping the cavity at two frequencies, each locked to a cavity mode. The cavity length in this case is set to ensure that the frequency located between the two SBS gain peaks coincides with one of the cavity modes. We found that the transfer function of the cavity in the vicinity of this mode exhibits a slightly broadened linewidth. The theoretical analysis confirmed that the WLC effect under the experimental parameters is indeed very small, and that it consistent with our experimental observations.

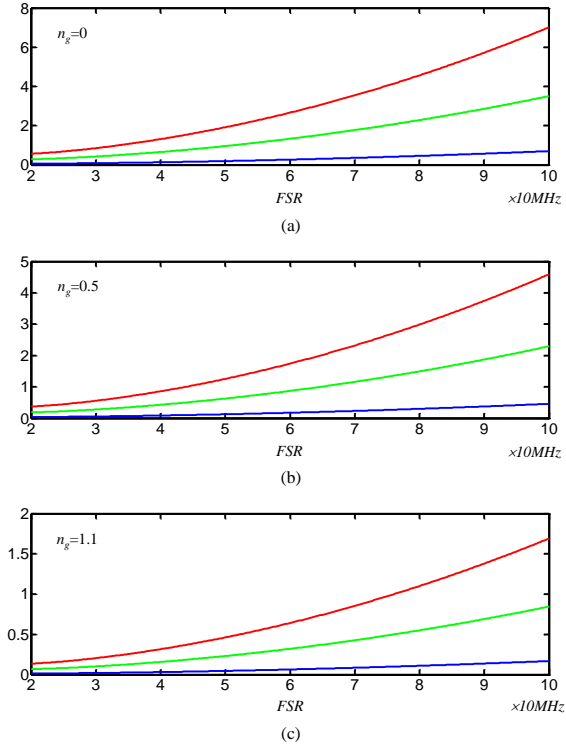

Fig. 9. (a) Cavity response measured for SBS absorption and gain by varying the pump power. (b) Numerical simulations for the cavity response corresponding to the experimental results. The colors of the graphs are the same as in (a). $G$ values used for the simulation are: $G=0.0002$ for $n_g=1.4538$ (yellow); $G=0.0003$ for $n_g=1.4557$ (cyan); $G=0.0005$ for $n_g=1.4595$ (magenta); $G=-0.0005$ for $n_g=1.4405$ (blue); $G=-0.0015$ for $n_g=1.4214$ (green); $G=-0.0018$ for $n_g=1.4156$ (red). The negative (positive) $G$ corresponds to SBS absorption (gain). (c) Calculated cavity response: (i) without SBS gain or absorption (black); (ii) on the WLC condition corresponding to $n_g=0$, $G=0.0759$ (red); (iii) non-dispersive cavity (i.e. $n_g=n_f$, $n_f=1.45$ for SMF-28e) including only SBS absorption-induced cavity loss.

In order to identify the conditions for observing a more prominent WLC effect, we studied theoretically the relations between the SBS pump, the cavity FSR and roundtrip loss needed for the realization of various group indices. We found a set of design rules which provide the necessary experimental parameters for realizing a WLC using SBS in fibers.

One of the important conclusions is that relatively high pump power is needed for attain zero group index in conventional fibers. In order to mitigate this limitation we proposed a new experimental scheme incorporating a non-resonant pumping scheme and special fibers having larger Brillouin coefficient than that of the conventional silica fiber. Another possibility is to incorporate optical amplification in the cavity to overcome the relatively high losses of such special fiber. Although the practical realization of WLC using SBS in fiber is not simple, the experimental studies and the accompanying analysis presented here show that it is feasible if the employed cavities are sufficiently long and the fiber propagation losses are sufficiently low.

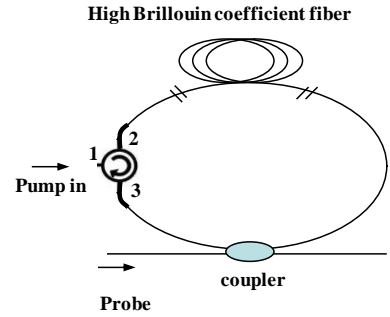

Fig. 10. Schematics of the proposed setup for WLC realization in fiber optics utilizing SBS